\documentclass[twocolumn,superscriptaddress, showkeys,showpacs,preprintnumbers,aps,pra]{revtex4}
\usepackage{graphicx}
\usepackage{dcolumn}
\usepackage{amssymb}
\usepackage{amsmath}
\usepackage{bbm}
\usepackage{setspace}
\usepackage{bbm}
\usepackage{hyperref}
\usepackage{CJK}
\usepackage{float}
\usepackage{url}
\hypersetup
{
	colorlinks=true, 
	citecolor=blue
}

\newcommand{\ii}{{\it i}}
\newcommand{\dd}{{\rm d}}

\newcommand{\sinc}{{\rm sinc}}
\newcommand{\erfi}{{\rm Erfi}}

\hypersetup
{
colorlinks=true, 
citecolor=blue
}

\begin{document}
\begin{CJK*}{UTF8}{gbsn}

\title{Non-perturbative dynamics of flat-band systems with correlated disorder }

\author{Qi Li(李骐)}
\affiliation{GBA Branch of Aerospace Information Research Institute, Chinese Academy of Sciences, Guangzhou 510535, China}
\affiliation{Guangdong Provincial Key Laboratory of Terahertz Quantum Electromagnetics, Guangzhou 510700, China}

\author{Junfeng Liu(刘军丰)}
\affiliation{School of Physics and Materials Science, Guangzhou University, Guangzhou 510006, China}

\author{Ke Liu(刘克)}
\affiliation{GBA Branch of Aerospace Information Research Institute, Chinese Academy of Sciences, Guangzhou 510535, China}
\affiliation{Guangdong Provincial Key Laboratory of Terahertz Quantum Electromagnetics, Guangzhou 510700, China}

\author{Zi-Xiang Hu(胡自翔)}
\affiliation{Department of Physics and Chongqing Key Laboratory for Strongly Coupled Physics, Chongqing University, Chongqing 401331, China}

\author{Zhou Li(李舟)}
\email{liz@aircas.ac.cn}
\affiliation{GBA Branch of Aerospace Information Research Institute, Chinese Academy of Sciences, Guangzhou 510535, China}
\affiliation{Guangdong Provincial Key Laboratory of Terahertz Quantum Electromagnetics, Guangzhou 510700, China}
\affiliation{University of Chinese Academy of Sciences, Beijing 100039, China}

\begin{abstract}
We develop a numerical method for the time evolution of Gaussian wave packets on flat-band lattices in the presence of correlated disorder. To achieve this, we introduce a method to generate random on-site energies with prescribed correlations. We verify this method with a one-dimensional (1D) cross-stitch model, and find good agreement with analytical results obtained from the disorder-dressed evolution equations. This allows us to reproduce previous findings, that disorder can mobilize 1D flat-band states which would otherwise remain localized. As explained by the corresponding disorder-dressed evolution equations, such mobilization requires an asymmetric disorder-induced coupling to dispersive bands, a condition that is generically not fulfilled when the flat-band is resonant with the dispersive bands at a Dirac point-like crossing. We exemplify this with the 1D Lieb lattice. While analytical expressions are not available for the two-dimensional (2D) system due to its complexity, we extend the numerical method to the 2D $\alpha-T_3$ model, and find that the initial flat-band wave packet preserves its localization when $\alpha = 0$, regardless of disorder and intersections. However, when $\alpha\neq 0$, the wave packet shifts in real space. We interpret this as a Berry phase controlled, disorder-induced wave-packet mobilization. In addition, we present density functional theory calculations of candidate materials, specifically $\rm Hg_{1-x}Cd_xTe$. The flat-band emerges near the $\Gamma$ point ($\bf{k}=$0) in the Brillouin zone.
\end{abstract}


\date{\today}

\maketitle

\textit{Introduction.}---Flat-band systems are characterized by completely dispersionless single-particle energy states, with the effective mass diverging. These systems were first introduced by Lieb in 1989 through two theorems on the Hubbard model \cite{Lieb}, where flat-band ferromagnetism was discovered in the repulsive Hubbard model with an unequal number of sites in the bipartite lattice \cite{Zsolt}. Flat bands have also been observed in certain decorated lattices \cite{Sutherland,Vidal,Tasaki, Mielke} with compact localized states for single electrons, as well as in sawtooth lattices that exhibit nonlinear localized modes \cite{Johansson}.

The Holstein model \cite{Zhou} exhibits a nearly flat band near the adiabatic limit, where the effective mass is very large but does not diverge. Consequently, the quasiparticle behavior in this model is described by Fermi liquid theory rather than Luttinger liquid theory. Certain short-range tight binding models display topological flat bands with a nonzero Chern number, providing insights into the fractional quantum Hall effect \cite{Sun, WangYF, Liu, XuPRX, Park}.

An intriguing lattice model called the $\alpha-T_3$ model exists in two dimensions (2D), which interpolates between the well-known graphene honeycomb lattice ($\alpha = 0$) and the dice lattice ($\alpha = 1$). The latter is realized in a tri-layer arrangement of cubic lattices along the (111) direction, as observed in materials such as $\rm SrTiO_3/SrIrO_3/SrTiO_3$ \cite{Fwang}. Moreover, an intermediate value of $\alpha = 1/\sqrt{3}$ can be achieved through critical doping of $\rm Hg_{1-x}Cd_xTe$\cite{Xu}, leading to intriguing orbital magnetic responses \cite{Raoux,Malcolm}. The Berry phase \cite{Illes} in this model undergoes continuous tuning from $\pi$ to 0 as $\alpha$ varies from 0 to 1, while the energy band structure remains unchanged. Unlike topological insulators with particle-hole asymmetry \cite{Zhou2}, where the Berry phase transitions from 0 to $\pi$ between non-relativistic Schrödinger and relativistic Dirac regimes in a discontinuous manner, the $\alpha-T_3$ model smoothly connects these two regimes without a Schrödinger term. Recently, flat bands have been experimentally realized in twisted bi-layer structures, such as boron nitride, graphene, and indium selenide \cite{MacDonald, Cao2018a, Cao2018b,Wu, Xian, Miao, Wei}, offering promising opportunities for integrating high-quality twisted bi-layer flat-band materials into waveguides or cavities for optoelectronic prototype devices. Theoretically,  catalogue of the naturally occuring three-dimensional stoichiometric materials with flat-band and the materials flat-band database website are also completed \cite{Regnault2022, calugaru2022}. 

In the field of photonic topological insulators, such as photonic Floquet topological insulators \cite{Szameit}, which was inspired by Haldane's work on designing directional optical waveguides in photonic crystals \cite{Haldane}, flat edge states emerge as a consequence of the system's topological properties. Recent advancements in experimental techniques have enabled the creation of flat bands by manipulating exciton-polaritons in arbitrary lattices \cite{Yamamoto} or through arrays of evanescently coupled optical waveguides \cite{Rodrigo, Sebabrata}. In quasi one-dimensional (1D) photonic systems, flat bands have been generated in experiments involving photonic Aharonov-Bohm cages, where the introduction of correlated disorder led to the observation of inverse Anderson localization \cite{Gligoric, Longhi, Kremer, Mukherjee, hang, Romer}.

In this study, we investigate the mobilization of flat-band states in both 1D and 2D systems in the presence of correlated disorder. It is worth noting that such disorder typically manifests in photons rather than in materials. By introducing correlated disorder into flat-band materials, we can explore intriguing dynamics of Gaussian wave packets. 

\begin{figure}[htbp]
    \centering    \includegraphics[width = 8cm]{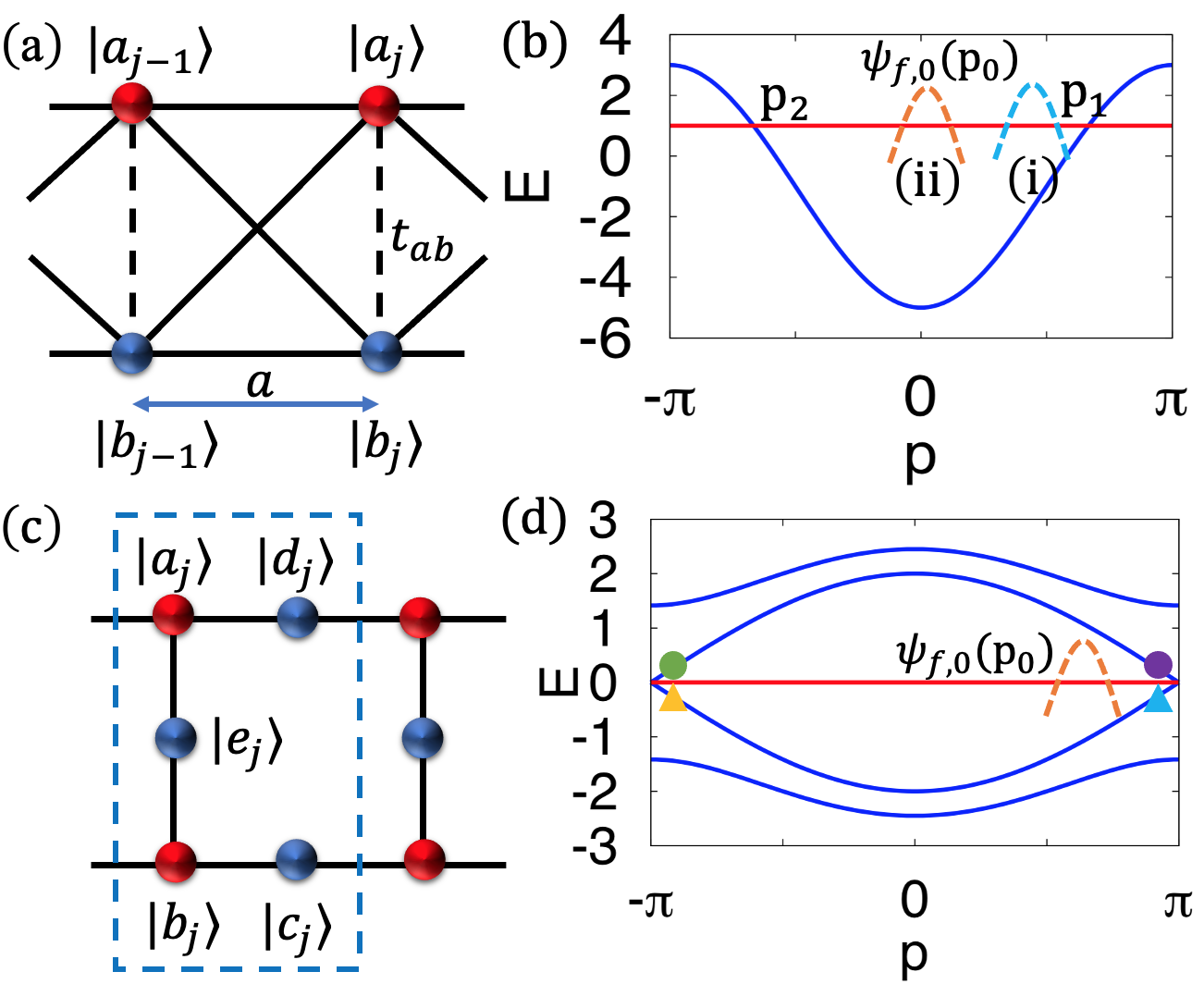}
    \caption{(Color online) (a) 1D cross-stitch lattice and (b) the band structure. $t_{ab}$ is the intra-cell hopping strength. An initial Gaussian wave packet located in (i) or (ii). (c,d) 1D Lieb lattice and the band structure, the flat band is in resonance with two inner dispersive bands, similar to the 2D $\alpha-T_3$ model. Four intersction points at $p=\pm \pi$ marked by purple, green dots for the upper band, and yellow, blue triangles for the lower band.}
\label{fig:lattice_sketch}
\end{figure}

\textit{1D model.}---The flat band emerges in the 1D cross-stitch lattice and Lieb lattice, depicted in Fig.~\ref{fig:lattice_sketch}(a) and (c) respectively. The energy bands in (b) are $E_f = t_{ab}$ (flat) and $E_d(p) = -4J\cos(p)-t_{ab}$ (dispersive), where $p$ represents the momentum (also in supplementary material).  The intersection points are determined by $p_{1,2}=\pm\arccos(-t_{ab}/2J)$, as illustrated in Fig.~\ref{fig:lattice_sketch}(b). The 1D Lieb lattice exhibits five bands with only two inner dispersive bands cross the central flat-band at $p = \pm \pi$, i.e., Dirac point-like crossing. 

\textit{Correlated disorder.}---To generate random onsite energy $V_\varepsilon^{\sigma}(j)$ ($\sigma\in\{a,b\}$, $\varepsilon$ labels different disorder realizations) satisfying a prescribed correlation function $ C_{\sigma \sigma}(j-j') =  \int  d\varepsilon p_{\varepsilon} V_\varepsilon^{\sigma}(j) V_\varepsilon^{\sigma}(j') = \int dq e^{iq(j-j')/\hbar}G_0(q)$, we employ the convolution method~\cite{Izrailev}. The modulation function $M(x)$, which is the Fourier transform of $[G_0(q)]^{1/2}$, is given by
\begin{equation}
M(x) = \int_{-\infty}^{\infty} [G_0(q)]^{\frac{1}{2}} e^{iqx} dq.
\end{equation}
The correlated onsite energy $V_\varepsilon^{a,b}(x)$ in real space is obtained by convolving the delta-correlated random number $\alpha(x)$ with the modulation function $M(x)$, and can be expressed as:
\begin{equation}
V_\epsilon^{a,b}(x) = \int_{-\infty}^{\infty} dx' \alpha(x-x') M(x')
\end{equation}
with $\alpha(x)$ satisfying the standard properties $\langle \alpha(x) \rangle = 0$ and $\langle \alpha(x) \alpha(x') \rangle = \delta(x-x')$.

The intra-sublattice correlations are the same, i.e. $G_{aa}(q)=G_{bb}(q)\equiv G_0(q)$. The inter-sublattice correlations differ by a pre-factor $\delta_{ab}$, so $G_{ab}(q) = G_{ba}(q) \equiv G_1(q) = \delta_{ab}G_{0}(q)$. Here, $\delta_{ab}=0,+1,-1$ corresponds to uncorrelated, fully correlated, and fully anti-correlated case, respectively. For Gaussian $C_{\sigma\sigma}(x)=C_0 e^{-(x/\ell)^2}$, we have $G_0(q) = \frac{C_0 \ell}{2\sqrt{\pi}\hbar} \exp\left[-\frac{1}{4}\left(\frac{q\ell}{\hbar}\right)^2\right]$, with the correlation length $\ell$.

\textit{Disorder-dressed evolution}---At short times, the ensemble-averaged dynamics of disordered models can be accurately described by the disorder-dressed evolution equation, where the incoherent nature primarily arises from the second-order term, known as the Lindblad term~\cite{GneitingPRA, GneitingPRB2020,GneitingPRB}. 
\begin{eqnarray}
\label{eq:master}
    \partial_t \overline{\rho}(t) &=& -\frac{i}{\hbar} \left[ \hat{H}_{\rm eff}, \overline{\rho}(t)  \right] + \sum_{\alpha \in \{\pm 1\}} \frac{2\alpha}{\hbar^2} \int_{-\infty}^{\infty} dq  \nonumber \\
    & \times & \sum_{\beta \in \{-1,0,1\} } \tilde{G}_\beta (q) \int_0^t d t' \mathcal{L}\left(\hat{L}_{q,\beta}^{(\alpha)}(t'),\overline{\rho}(t)\right)
\end{eqnarray}
where $\mathcal{L}(\hat{L},\rho)=\hat{L}\rho \hat{L}^\dagger - \frac{1}{2}\hat{L}^\dagger\hat{L}\rho -\frac{1}{2}\rho \hat{L}^\dagger\hat{L} $ and $\tilde{G}_{0} (q)  = G_{0}(q)(1+\delta_{ab})/2, \tilde{G}_{1} (q) =\tilde{G}_{-1} (q)   = G_{0}(q)(1-\delta_{ab})/4$. The effective Hamiltonian $\hat{H}_{\rm eff}(t) $ and the Lindblad operators $\hat{L}_{q,\beta}^{(\alpha)}(t)$ are given  in~\cite{supplementary}.

Next, we evaluate Eq.~(\ref{eq:master}) for the 1D cross-stitch model and present partial new results. We treat the dispersive band linearly~\cite{GneitingPRB} near the intersections:
\begin{equation} 
\label{eq:Hbar}
\hat{H} = v(\hat{p}-p_1)\otimes \vert d \rangle \langle d \vert
\end{equation} 
where the velocity $v$ denotes the slope of the dispersive band at the intersections. Assuming the dispersive band state component $\overline{\rho}_d$ remains negligible during the time scales of interest, we can derive a closed evolution equation for the flat-band component $\overline{\rho}_f$~\cite{supplementary}:
\begin{align} \label{eq:master_equation}
\partial_t \overline{\rho}_f(p) =&-\sum_{j=1,2} \Gamma^{(j)}_t(p-p_j)\overline{\rho}_f(p) \\
 &+\frac{t(1+\delta_{ab})}{\hbar^2}  \int_{-\infty}^{\infty} \dd q G_0(q)\Big[\overline{\rho}_f(p-q)-\overline{\rho}_f(p)\Big] \nonumber
\end{align}
where $\Gamma_t(p)=\frac{1-\delta_{ab}}{\hbar^2}  \int_{-\infty}^{\infty} \dd q {G}_0(q)t\,\sinc[\frac{vt(q-p)}{\hbar}]$. The first term in Eq.~\eqref{eq:master_equation} describes the decay into the dispersive band, while the second term describes the dephasing process that gradually broadens the momentum distribution of the flat-band state ~\cite{GneitingPRB, supplementary}.
Note that $\sinc[\frac{vt(q-p)}{\hbar}]= \int_{-1/2}^{1/2} \dd x \exp[\frac{2\ii vt(q-p)x}{\hbar}]$, and after some algebra we obtain
\begin{align} \label{Eq:gamma1}
\Gamma_t(p)=&\frac{1-\delta_{ab}}{2\hbar}\frac{\pi\ii}{v}G_0(p) \nonumber\\ &\times\Bigg[\erfi\Big[\frac{p\ell}{2\hbar}-\frac{\ii tv}{\ell}\Big]-\erfi\Big[\frac{p\ell}{2\hbar}+\frac{\ii tv}{\ell}\Big]\Bigg]
\end{align}
where $\erfi$ is the imaginary error function. Eq.~(\ref{Eq:gamma1}) is not explicitly given in~\cite{GneitingPRB}. Moreover, we present the evolution of $\overline{\rho}_d$ in~\cite{supplementary}.

\textit{Decay and dephasing.}--- The system is diagonalized under periodic boundary conditions with $N=100$ unit cells, and an average is taken over $K=200$ disorder realizations. The Gaussian wave packet on the flat-band is $\psi_0(x) = \frac{1}{\sqrt{\sqrt{\pi} \sigma}} \exp\left[-\frac{(x-x_0)^2}{2\xi^2}+ip_0x\right]$, where $x_0 = 50a$ (with lattice constant $a$) denotes the initial position , $\xi^2 = 12a^2$ represents the position uncertainty, and $p_0$ is the initial average momentum. 
For each disorder realization, the time evolution is determined through $\vert \psi(t) \rangle = \sum_n \exp(-i\varepsilon_n t) \langle \psi_n \vert \psi_0 \rangle \vert \psi_n \rangle$, with the eigenvalues $\varepsilon_n$ and the eigenvectors $\psi_n$ of the Hamiltonian $\hat{H}_\varepsilon$.
\begin{figure}[htbp]
    \centering    \includegraphics[width = 8.5cm]{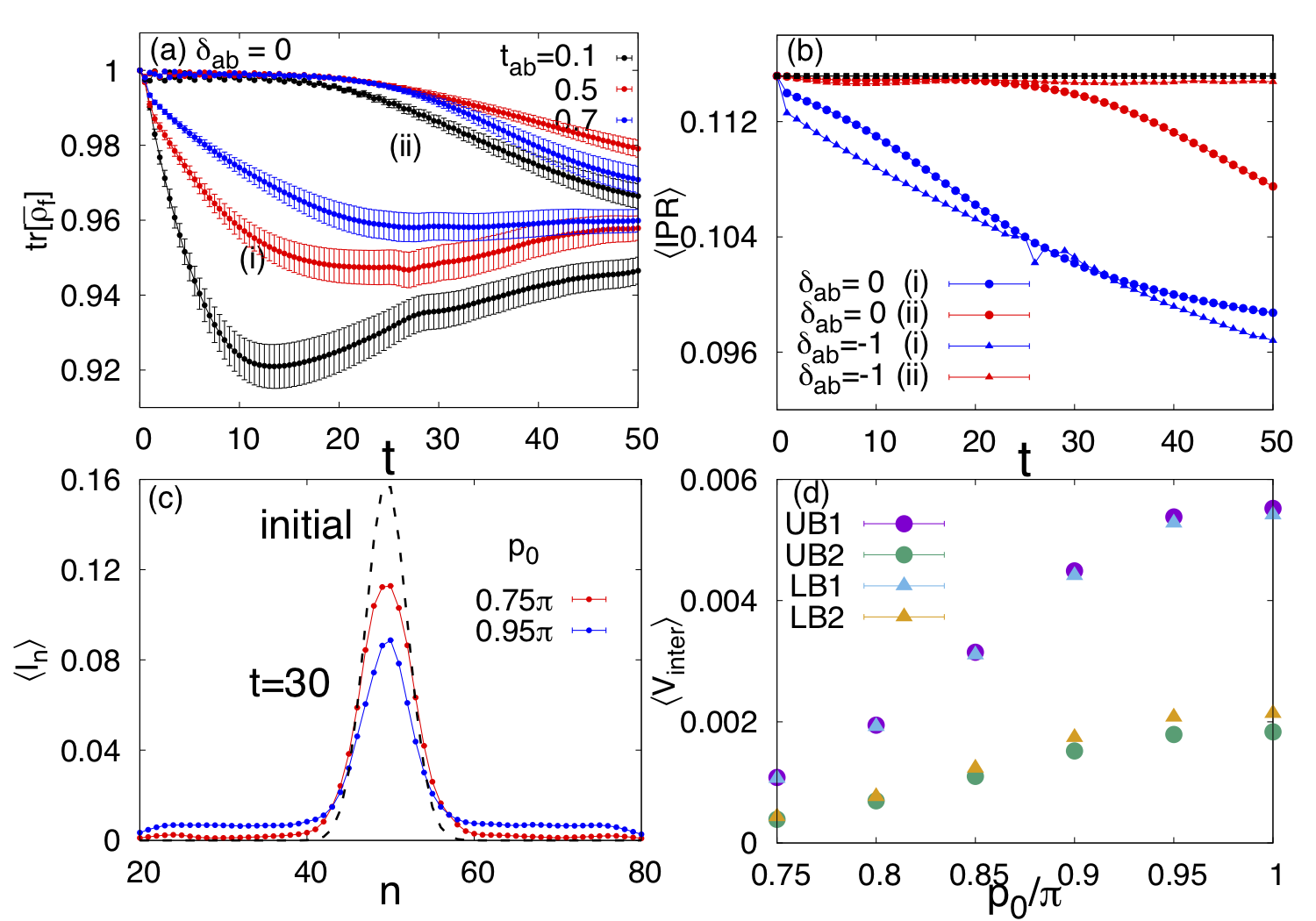}
    \caption{(Color online)Disorder-induced flat-band dynamics in (a,b)1D cross-stitch lattice and (c,d)1D Lieb lattice. Tune the location of the flat-band through varying $t_{ab}$, an interference effect between two intersections is indicated from the reversed order in the decay behavior for case (ii). (b)The ensemble averaged inverse participation ratio  $\langle \rm IPR \rangle$. Black dot-dashed for $\delta_{ab}=1$. (c)The intensity profile $\langle \rm I_n \rangle$ at t = 30 for different momentum space locations $p_0$. (d)Disorder-averaged inter-band coupling strength $\langle V_{inter} \rangle = \langle \psi_d \vert \hat{V_{\epsilon}} \vert \psi_{f,0}(p_{0}) \rangle$ between initial flat-band and upper(lower) dispersive bands (UB,LB). Labels are the same as in Fig.~\ref{fig:lattice_sketch}(d). Standard errors are used as error bars in each figure.}
    \label{fig:rhof_trace}
\end{figure}

In Fig.~\ref{fig:rhof_trace}(a) we present interference patterns goes beyond Eq.~\eqref{eq:master_equation}, which treats the two intersections independently. The (i) resonant or (ii) detuned case depends on the location of the initial flat-band state, as depicted in Fig~\ref{fig:lattice_sketch}(b). The decayed state component can propagate in the dispersive band and couple back into the flat-band, resulting in the effective diffusion of the flat-band state. As shown
in Fig.~\ref{fig:rhof_trace}(b), the inverse participation ratio ${\rm IPR}(t) = \sum_j I_j^2(t) / \sum_j I_j(t)$ is analyzed to quantify the distribution of the flat-band wave packet, where the total intensity in each unit cell is defined as $I_j(t) = \vert a_j(t)\vert^2+\vert b_j(t) \vert^2$. Consistent with previous results, the trivial decoupled case with $\delta_{ab} = 1$ shows no diffusion. For $\delta_{ab} = 0$ or $-1$, the detuned case (ii) exhibits significantly modified diffusion magnitudes compared to the resonant case (i). The difusion is directional if there is a preferential coupling to one of the dispersive band \cite{supplementary}. 

For the 1D Lieb lattice we have two dispersive bands at the same intersection $p=\pi,-\pi$, however, the upper or lower dispersive bands contain both $v$ and $-v$ velocities, the wave packet on the flat band does not move towards a specific direction, as shown in Fig.~\ref{fig:rhof_trace} (c). The inter-band coupling strength to the upper and lower dispersive bands are different, as in Fig.~\ref{fig:rhof_trace} (d). 

\begin{figure}[htbp]
    \centering    \includegraphics[width =8.5cm]{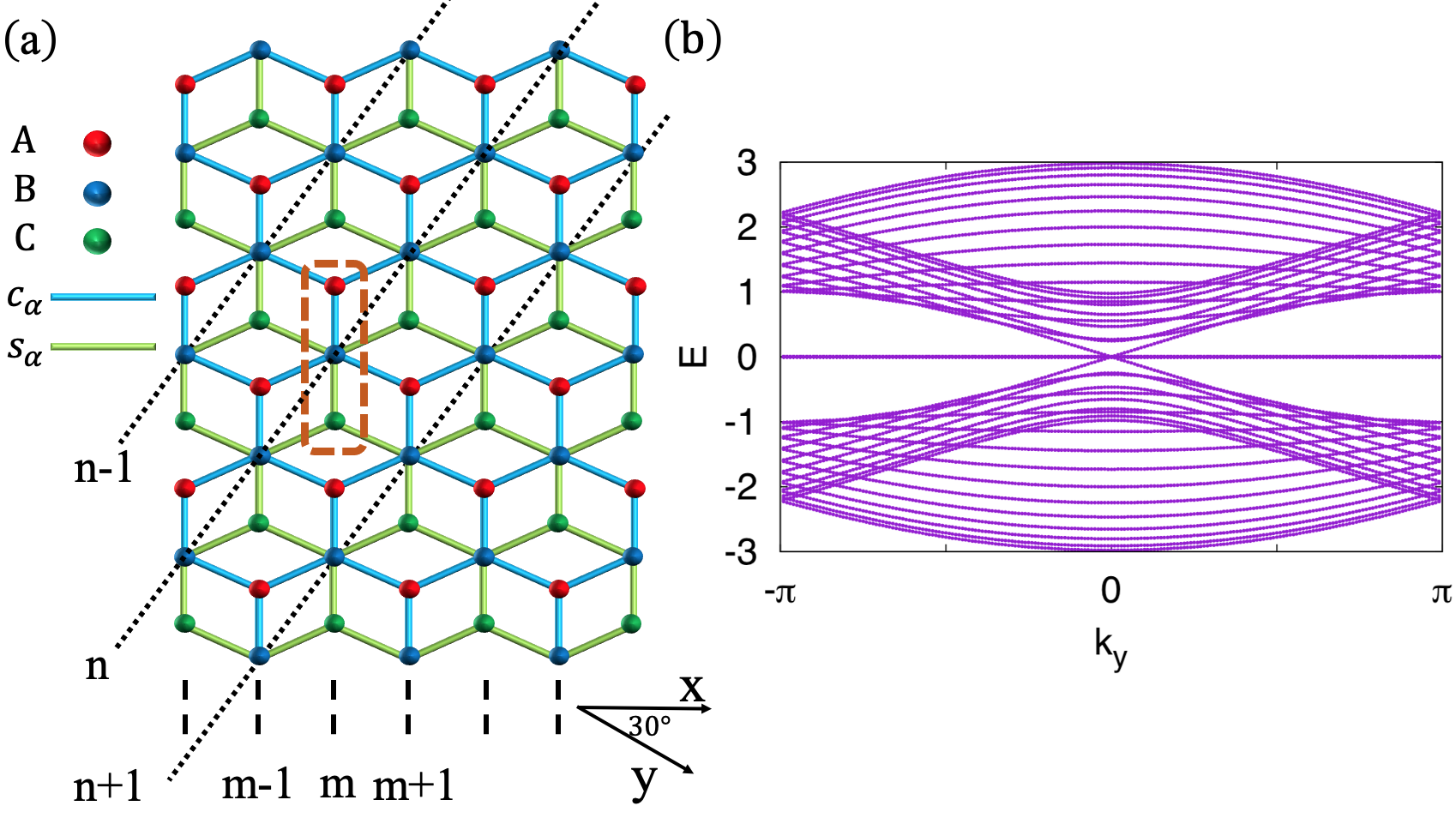}
    \caption{(Color online) (a) An armchair nanotube structure from  $\alpha-T_3$ lattice. Only nearest-neighbor hopping are considered. Blue links represent the hopping between $A$ and $B$ sites with amplitude $C_\alpha = \frac{1}{\sqrt{1+\alpha^2}} $ (in unit of $J$), and green links for the hopping between $B$ and $C$ atoms with amplitude $S_\alpha = \frac{\alpha}{\sqrt{1+\alpha^2}}$ such that $C_\alpha^2 + S_\alpha^2 =1$. The location of unit cell is labelled by $m$, $n$ in the real space. (b)Band structure of armchair nanotube with width $N_x = 20$. For a fixed $k_y$   the flat-band degeneracy is $N_x$-fold. The Dirac point is on $(\pm \frac{4\pi}{3\sqrt{3}a},0)$. Note that the spectrum does not dependent on $\alpha$.}
    \label{fig:alphat3-lattice}
\end{figure}

\textit{2D model.}--- The $\alpha-T_3$ model is a triangular Bravais lattice with three sites per unit cell labeled as $A$, $B$, and $C$, as shown in Fig.~\ref{fig:alphat3-lattice} (a). The hopping amplitudes between sites are given by $C_\alpha J$ for hopping between $A$ and $B$, and $S_\alpha J$ for hopping between $B$ and $C$. Here, $J$ represents the hopping strength and $\alpha$ is a parameter that determines the ratio between the two hopping amplitudes. In the continuum limit, with no on-site potential, the $\alpha-T_3$ model exhibits one flat band with energy $E_0 = 0$ and two dispersive bands $E_{\pm}$ regardless of the value of $\alpha$ \cite{supplementary}. It is worth noting that varying $\alpha$ from $0$ to $1$ allows for a continuous change in the Berry phase. Fig.~\ref{fig:alphat3-lattice} (b) gives the band structure of $\alpha-T_3$ armchair nanotube with flat-band crossing other bands at $k_y = 0$. In the site basis $\Psi_{m,n} = (\Psi_{m,n}^A,\Psi_{m,n}^B, \Psi_{m,n}^C)^T$, where $m$ and $n$ represent the lattice indices with armchair edges in the $x$ direction and periodic boundary conditions in the $y$ direction, the tight-binding model for the nanotube can be written as ($J = 1$ for simplicity):~\cite{Kohmoto, Chen,Oriekhov}:
\begin{eqnarray}
\label{eq:tb-ribbon}
&& -C_\alpha \Psi_{m,n}^B-C_\alpha \Psi_{m-1,n-1}^B-C_\alpha \Psi_{m+1,n}^B = E\Psi_{m,n}^A  ,\nonumber \\
&& -C_\alpha \Psi_{m,n}^A-C_\alpha \Psi_{m-1,n}^A-C_\alpha \Psi_{m+1,n+1}^A  \nonumber \\
&& -S_\alpha \Psi_{m,n}^C-S_\alpha \Psi_{m-1,n-1}^C-S_\alpha \Psi_{m+1,n}^C = E\Psi_{m,n}^B  , \nonumber \\
&& -S_\alpha \Psi_{m,n}^B-S_\alpha \Psi_{m-1,n}^B-S_\alpha \Psi_{m+1,n+1}^B = E\Psi_{m,n}^C \nonumber \\
\end{eqnarray}
\begin{figure}[htbp]
    \centering
    \includegraphics[width=8.5cm]{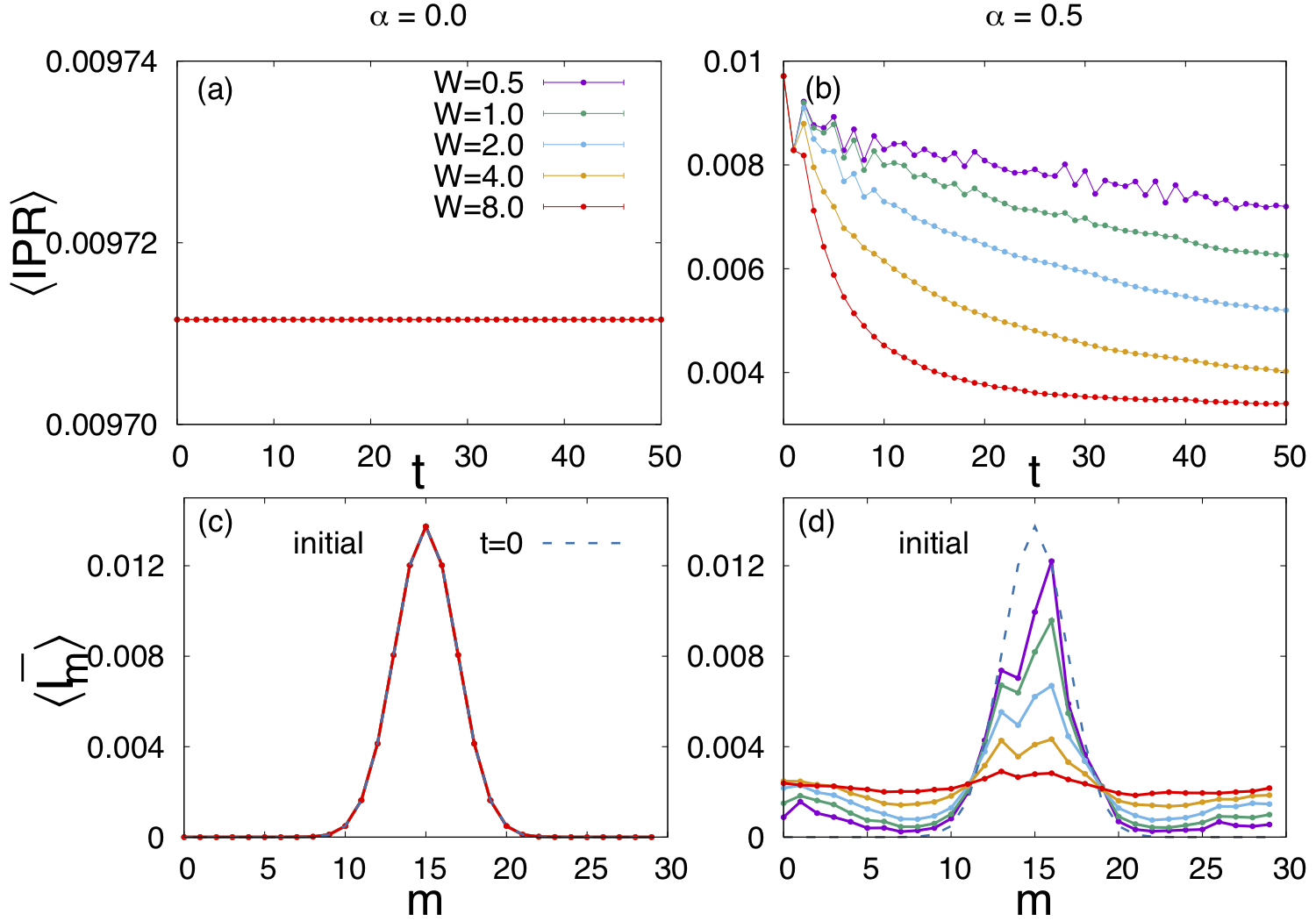}
    \caption{(Color online) (a-b) The ensemble average $\langle \rm IPR \rangle$ versus time $t$. Time evolution is involved in $N_x \times N_y$ armchair nanotube system with $N_x = 30, N_y =15$, averaging over $K = 100$ disorder realizations. Disorder strength $W$ ranges from $0.5-8.0$ and the initial average momentum in x direction is set as $p_{0x} = 0$. (c-d) Disorder-averaged mean intensity profile $\langle \rm \overline{I_m} \rangle$ at $t=50$. Dashed line represents the initial mean intensity profile. The left and right columns corresponds to $\alpha = 0, 0.5$. }
    \label{fig:IPR}
\end{figure}

In the presence of correlated disorder $V^\sigma_\varepsilon(\mathbf{r})$ with only intra-sublattice correlations $C_{\sigma\sigma}(\mathbf{r}) = C_0 e^{-(|\mathbf{r}-\mathbf{r'}|^2/\ell^2)}$,  with fixed correlation length $\ell = 6a$, where $\sigma \in \rm \{A,B,C \}$, the dynamics of the isolated compact flat-band mode in the $\alpha-T_3$ nanotube with finite size $N_x \times N_y$ can be investigated. The initial state is prepared as a flat-band state with a Gaussian prefactor in the $x$ direction, given by $\psi_{f,0} \sim \exp\left[-\frac{(x-x_0)^2}{2\xi^2}+ i p_{0x} x \right] \Psi_f$, where $\Psi_f = (S_\alpha, 0, -C_\alpha)^T$ and $x_0 = N_x/2$. In $y$ direction, the distribution is uniform. This initial state is localized on either the $A$ or $C$ sites, depending on the value of $\alpha$. The dynamics of this wave packet are studied in the presence of disorder.

Fig.~\ref{fig:IPR}(a) and (b) illustrate the disorder-averaged inverse participation ratio ($\langle \text{IPR} \rangle$) and its dependence on the disorder strength $W$. When $\alpha = 0$, the flat-band states $\psi_{f,0}$ are trapped on isolated $C$ sites in the middle of the honeycomb lattice, with no hoppings to the other sites. The numerically calculated IPR confirms the presence of localization in this case, as shown in Fig.~\ref{fig:IPR}(a).

In contrast, when $\alpha \neq 0$, the flat-band states are initially distributed on both $A$ and $C$ sites. The presence of correlated disorder induces anti-localization in the flat-band mode. As the disorder strength $W$ increases, a larger number of sites are populated by the wave packet, as observed in Fig.~\ref{fig:IPR}(b). It is worth noting that larger values of $\alpha$ can enhance the rate of diffusion gradually. For the 1D cross-stitch model with two intersections, the wave packet moves when it is closer to one of the intersections $p_1$ or $p_2$. For 2D $\alpha-T_3$ armchair ribbon, the wave function $\Psi=e^{ik_yy}\Psi(x)$, where $\Psi(x)\propto e^{ik_xx}-e^{-ik_xx}$ from the armchair boundary condition. Because of the different symmetry, the wave packet moves in the transverse direction $x$. This is not seen in the 1D Lieb lattice. We verify that this mobilization in 2D $\alpha-T_3$ armchair ribbon happens for both uncorrelated and correlated disorder.

To analyze the spreading of the wave packet, the mean intensity $\overline{I_m} = \sum_n I_{mn} / N_y$ is calculated as a function of the lattice index $m$ in the $x$ direction after a propagation time of $t = 50$. The $\alpha = 0$ case preserves its initial intensity profile, as shown in Fig.\ref{fig:IPR}(c). However, in the case of $\alpha \neq 0$, the breakdown of compact localization is observed, and the wave packet spreads over a larger number of lattice sites, as seen in Fig.\ref{fig:IPR}(d). Stronger disorder leads to wider spreading of the wave packet, but transitions to regular Anderson localization are not observed in this system. These results highlight the influence of correlated disorder on the dynamics and localization properties of the flat-band states in the $\alpha-T_3$ nanotube model.

\begin{figure}[htbp]
    \centering    \includegraphics[width=8.5cm]{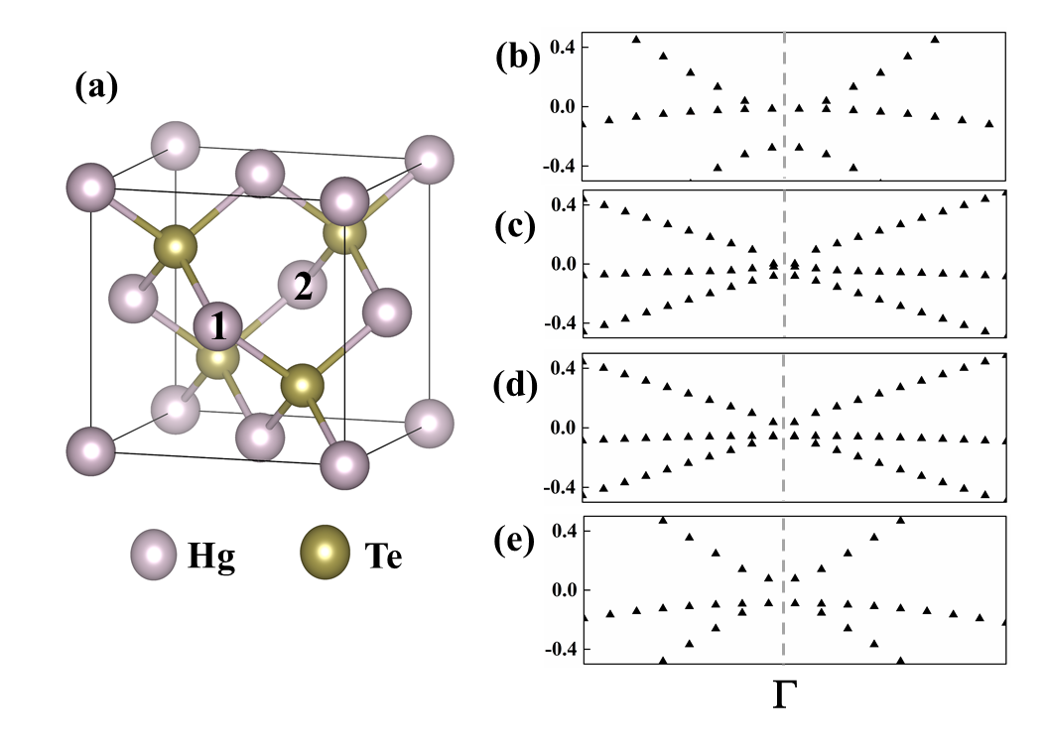}
    \caption{(a) Zinc blende structure. (b-e) Band structure of $\rm Hg_{1-x}Cd_xTe$ for $x$=0, 0.15625, 0.1875 and 0.25 respectively. For $x=0.25$, two Hg atoms in (a) labeled by 1 and 2 are replaced by two Cd atoms. For (c) and (d), 5 and 6 Hg atoms are replaced by Cd atoms in the 32-atom super-cell respectively.}
    \label{fig:band}
\end{figure}

\textit{Candidate material.}---The first principles method based on density functional theory (DFT) is used to calculate the electronic properties, employing the DS-PAW software on the Device Studios platform. DS-PAW is a software that utilizes the projection plus plane-wave (PAW) method~\cite{blochl1994}. The Perdew-Burke-Ernzerhof (PBE) exchange-correlation energy functional within the generalized gradient approximation (GGA) is employed~\cite{perdew1996} to optimize the atomic structure. The local density approximation (LDA) or generalized gradient approximation (GGA) tends to underestimate the band gap~\cite{aouail2021, guo2012}. Therefore, we use the hybrid functional theory (HSE06) to calculate the band structure. The atomic positions are fully optimized until the force on each atom is less than 0.05eV/$\AA$, and the electron iteration setting is less than $10^{-5}$eV for convergence. The kinetic energy cutoff of the wave function in the plane wave is set to 400 eV. Brillouin zone integrations are obtained using a k-mesh sampling mesh generated according to the Gamma-centered method. The band structure of $\rm Hg_{1-x}Cd_xTe$ is calculated for $x=0$ and 0.25, as shown in Fig.~\ref{fig:band}. The gap between two Dirac cones at the $\Gamma$ point closes and reopens, indicating that the band gap is expected to close at a critical value of $x_c$ with $0<x_c<0.25$.

\textit{Conclusion.}---In summary, our investigations on wave-packet dynamics in the presence of correlated disorder in both 1D and 2D systems have provided valuable insights into the behavior of flat-band states intersecting with dispersive bands. 

While analytical expressions using master equations are not available for the 2D system due to its complexity with multiple bands, we believe that our predictions and observations can be verified in future theoretical ~\cite{GneitingCredit} and experimental studies. The understanding gained from these investigations can contribute to the development of novel materials and systems with tailored flat-band properties and their potential applications in various fields. The study of investigating  dynamics of topological flat-bands with non-zero Chern number under the influence of disorders would be more interesting and provides a new platform  and direction for our future research.

\begin{acknowledgements}The work is supported by NSFC Grant No. 61988102, the Key Research and Development Program of Guangdong Province (2019B090917007) and the Science and Technology Planning Project of Guangdong Province (2019B090909011). Q. L. acknowledges Guangzhou Basic and Applied Basic Research project No. 2023A04J0018. Z. L. acknowledges the support of funding from Chinese Academy of Science E1Z1D10200 and E2Z2D10200; from ZJ project 2021QN02X159 and from JSPS Grant No. PE14052
and P16027. We gratefully acknowledge HZWTECH for providing computation facilities. Z.-X. H. was supported by NSFC Grant No. 11974064, No.12147102, and the Fundamental Research Funds for the Central Universities under Grant No. 2020CDJQY-Z003.
\end{acknowledgements}

\end{CJK*}

\begin{thebibliography}{99}
\bibitem{Lieb}
Lieb E H
\href{https://doi.org/10.1103/PhysRevLett.62.1201}
{1989 \textit{Phys. Rev. Lett.} \textbf{62}, 1201}

\bibitem{Zsolt}
Gul\'acsi Z, Kampf A and Vollhardt D
\href{https://doi.org/10.1103/PhysRevLett.99.026404}
{2007 \textit{Phys. Rev. Lett.} \textbf{99}, 026404}

\bibitem{Sutherland}
Sutherland B
\href{https://doi.org/10.1103/PhysRevB.34.5208}
{1986 \textit{Phys. Rev. B} \textbf{34}, 5208}

\bibitem{Vidal}
Vidal J, Mossen R and Dou\ifmmode \mbox{\c{c}}\else \c{c}\fi{}ot B
\href{https://doi.org/10.1103/PhysRevLett.81.5888}
{1998 \textit{Phys.Rev.Lett.} \textbf{81}, 5888}

\bibitem{Tasaki}
Tasaki H
\href{https://doi.org/10.1103/PhysRevLett.69.1608}
{1992 \textit{Phys. Rev. Lett.} \textbf{69}, 1608}

\bibitem{Mielke}
Mielke A and Tasaki H
\href{https://doi.org/10.1007/BF02108079}
{1993 \textit{Comm. Math. Phys.} \textbf{158}, 341}

\bibitem{Johansson}
Johansson M, Naether U and Vicencio R A
\href{https://doi.org/10.1103/PhysRevE.92.032912}
{2015 \textit{Phys. Rev. E} \textbf{92}, 032912}

\bibitem{Zhou}
 Li Z, Baillie D, Blois C and Marsiglio F
\href{https://doi.org/10.1103/PhysRevB.81.115114}
 {2010 \textit{Phys. Rev. B} \textbf{81}, 115114}
 
\bibitem{Sun}
Sun K, Gu Z C, Katsura H and Sarma S D
\href{https://doi.org/10.1103/PhysRevLett.106.236803}
{2011 \textit{Phys. Rev. Lett.} \textbf{106}, 236803}
 
\bibitem{WangYF}
Wang Y F, Gu Z C, Gong C D and Sheng D N
\href{https://doi.org/10.1103/PhysRevLett.107.146803}
{2011 \textit{Phys. Rev. Lett.} \textbf{107}, 146803}

\bibitem{Liu}
Liu Z and Bhatt R N
\href{https://doi.org/10.1103/PhysRevLett.117.206801}
{2016 \textit{Phys. Rev. Lett.} \textbf{117}, 206801}

\bibitem{XuPRX}
Xu F, Sun Z,  Jia T T,  Liu C,  Xu C,  Li C S , Gu Y, Watanabe K, Taniguchi T,  Tong B B, Jia J F,  Shi Z W, Jiang S W,  Zhang Y, Liu X X, and Li T X
\href{https://doi.org/10.1103/PhysRevX.13.031037}
{2023 \textit{Phys. Rev. X} \textbf{13}, 031037}

\bibitem{Park}
Park H, Cai J Q, Anderson E et al.
\href{https://doi.org/10.1038/s41586-023-06536-0}
{2023 \textit{Nature} \textbf{622}, 74-79}

\bibitem{Fwang}
Wang F and Ran Y
\href{https://doi.org/10.1103/PhysRevB.84.241103}
{2011 \textit{Phys. Rev. B} \textbf{84}, 241103(R)}

\bibitem{Xu}
Xia B W, Wang R, Chen Z J, Zhao Y J and Xu H
\href{https://doi.org/10.1103/PhysRevLett.123.065501}
{2019 \textit{Phys. Rev. Lett.} \textbf{123}, 065501}

\bibitem{Raoux}
Raoux A, Morigi M, Fuchs J N, Pi\'{e}chon P and Montambaux G
\href{https://doi.org/10.1103/PhysRevLett.112.026402}
{2014 \textit{Phys. Rev. Lett.} \textbf{112}, 026402}

\bibitem{Malcolm}
Malcolm J D and Nicol E J
\href{https://doi.org/10.1103/PhysRevB.92.035118}
{2015 \textit{Phys. Rev. B} \textbf{92}, 035118}

\bibitem{Illes}
Illes E, Carbotte J P and Nicol E J
\href{https://doi.org/10.1103/PhysRevB.92.245410}
{2015 \textit{Phys. Rev. B.} \textbf{92}, 245410}
  
\bibitem{Zhou2}
Li Z and Carbotte J P
\href{https://doi.org/10.1103/PhysRevB.89.085413}
{2014 \textit{Phys. Rev. B} \textbf{89}, 085413}

\bibitem{MacDonald}
Bistritzer R and MacDonald A H
\href{https://doi.org/10.1073/pnas.1108174108}
{2011 \textit{Proc. Natl. Acad. Sci.} \textbf{108}, 12233-12237} 

\bibitem{Cao2018a}
Cao Y, Fatemi V, Fang S, Watanable K, Tanigunchi T, Kaxiras E and Jarillo-Herrero P 
\href{https://doi.org/10.1038/nature26160}
{2018 \textit{Nature} \textbf{556}, 43-50}

\bibitem{Cao2018b}
Cao Y, Fatemi V, Demir A, Fang S, Tomarken S L, Luo J Y, Yamagishi J D S, Watanable K, Taniguchi T, Kaxiras E, Ashoori R C and Jarillo-Herrero P
\href{https://doi.org/10.1038/nature26154}
{2018 \textit{Nature} \textbf{556}, 80-84}

\bibitem{Wu} 
Huang G H, Xu Z F and Wu Z G,
\href{https://doi.org/10.1103/PhysRevLett.129.185301}
{2022 \textit{Phys. Rev. Lett.} \textbf{129}, 185301}
 
\bibitem{Xian} 
Xian L D, Kennes D M, Dejean N T, Altarelli M and Rubio A
\href{https://doi.org/10.1021/acs.nanolett.9b00986}
{2019 \textit{Nano Lett.} \textbf{19}, 8, 4934–4940}

\bibitem{Miao} 
Li Q, Cheng B, Chen M Y, Xie B, Xie Y Q, Wang P F, Chen F Q, Liu Z L, Watanabe K, Taniguchi T, Liang S J, Wang D, Wang C J, Wang Q H, Liu J P and Miao F
\href{https://doi.org/10.1038/s41586-022-05106-0}
{2022 \textit{Nature} \textbf{609}, 479–484}
 
\bibitem{Wei}
Tao S D, Zhang X L, Zhu J J, He P M, Yang S Y A, Lu Y H and Wei S H
\href{https://doi.org/10.1021/jacs.1c11953}
{2022 \textit{J. Am. Chem. Soc.} \textbf{144}, 9, 3949-3956}

\bibitem{Regnault2022}
Regnault N, Xu Y F, Li M R, Ma D S, Jovanovic M, Yazdani A, Parkin S S P, Felser C, Schoop L M, Ong N P, Cava R J, Elcoro L, Song Z D and Bernevig A
\href{https://doi.org/10.1038/s41586-022-04519-1}
{2022 \textit{Nature} \textbf{603}, 824-828}

\bibitem{calugaru2022}
C\v{a}lu\v{a}ru D, Chew A, Elcoro L, Xu Y F, Regnault N, Song Z D and Bernevig B A 
\href{https://doi.org/10.1038/s41567-021-01445-3}
{2022 \textit{Nat. Phys.} \textbf{18}, 185-189}
 
\bibitem{Szameit}
Rechtsman M C, Zeuner J M, Plotnik Y, Lumer Y, Podolsky D, Dreisow F, Nolte S, Segev M and Szameit A
\href{https://doi.org/10.1038/nature12066}
{2013 \textit{Nature} \textbf{496}, 196}

\bibitem{Haldane}
Haldane F D M and Raghu S
\href{https://doi.org/10.1103/PhysRevLett.100.013904}
{2008 \textit{Phys. Rev. Lett.} \textbf{100}, 013904}

\bibitem{Yamamoto}
Kim N Y, Kusudo K, Wu C, Masumoto N, L\"offler A, H\"ofling S, Kumada N, Worschech L, Forchel A and Yamamoto Y
\href{https://doi.org/10.1038/nphys2012}
{2011 \textit{Nature} \textbf{7}, 681-686}

\bibitem{Rodrigo}
Vicencio R A, Cantillano C, Morales-Inostroza L, Real B, Mej\'{\i}a-Cort\'es C, Weimann S, Szameit A and Molina M I,
\href{https://doi.org/10.1103/PhysRevLett.114.245503}
{2015 \textit{Phys. Rev. Lett.} \textbf{114}, 245503}

\bibitem{Sebabrata}
Mukherjee S, Spracklen A, Choudhury D, Goldman N, \"Ohberg P, Andersson E and Thomson R
\href{https://doi.org/10.1103/PhysRevLett.114.245504}
{2015 \textit{Phys. Rev. Lett.} \textbf{114}, 245504}

\bibitem{Gligoric}
Gligori\'c G, Leykam D and Maluckov A
\href{https://doi.org/10.1103/PhysRevA.101.023839}
{2020 \textit{Phys. Rev. A} \textbf{101}, 023839}

\bibitem{Longhi}
Longhi S
\href{https://doi.org/10.1364/OL.430196}
{2021 \textit{Opt. Lett.} \textbf{46}, 2872}

\bibitem{Kremer}
Kremer M, Petrides I, Meyer E, Heinrich M, Zilberberg O and Szameit A
\href{https://doi.org/10.1038/s41467-020-14692-4}
{2020 \textit{Nat. Commun.} \textbf{11}, 907}

\bibitem{Mukherjee}
Mukherjee S, Liberto M D, \"Ohberg P, Thomson R and Goldman N
\href{https://doi.org/10.1103/PhysRevLett.121.075502}
{2018 \textit{Phys. Rev. Lett.} \textbf{121}, 075502}

\bibitem{hang}
Li H, Dong Z L, Longhi S, Liang Q, Xie DZ and Yan B
\href{https://doi.org/10.1103/PhysRevLett.129.220403}
{2022 \textit{Phys. Rev. Lett.} \textbf{129}, 220403}

\bibitem{Romer} 
Liu J, Danieli C, Zhong J X and R\"omer R A
\href{https://doi.org/10.1103/PhysRevB.106.214204}
{2022 \textit{Phys. Rev. B} \textbf{106}, 214204}

\bibitem{Izrailev}
Izrailev F M, Krokhin A A and Makarov N M
\href{https://doi.org/10.1016/j.physrep.2011.11.002}
{2012 \textit{Phys. Rep.} \textbf{512}, 125}

\bibitem{GneitingPRA}
Gneiting C and Nori F
\href{https://doi.org/10.1103/PhysRevA.96.022135}
{2017 \textit{Phys. Rev. A} \textbf{96}, 022135}

\bibitem{GneitingPRB2020}
Gneiting C
\href{https://doi.org/10.1103/PhysRevB.101.214203}
{2020 \textit{Phys. Rev. B} \textbf{101}, 214203}

\bibitem{GneitingPRB}
Gneiting C, Li Z and Nori F
\href{https://doi.org/10.1103/PhysRevB.98.134203}
{2018 \textit{Phys. Rev. B} \textbf{98}, 134203}

\bibitem{supplementary}
See Supplemental Material at http://.....  for a detailed derivations and more numerical results. 

\bibitem{FlachRPL}
Bodyfelt J D, Leykam D, Danieli C, Yu X and Flach S
\href{https://doi.org/10.1103/PhysRevLett.113.236403}
{2014 \textit{Phys. Rev. Lett.} \textbf{113}, 236403}

\bibitem{FlachEPL}
Flach S, Leykam D, Bodyfelt J D, Matthies P and Desyatnikov A S
\href{https://doi.org/10.1209/0295-5075/105/30001}
{2014 \textit{Europhys. Lett.} \textbf{105}, 30001}

\bibitem{Kohmoto}
Kohmoto M and Hasegawa Y
\href{https://doi.org/10.1103/PhysRevB.76.205402}
{2007 \textit{Phys. Rev. B} \textbf{76}, 205402}

\bibitem{Chen}
Chen Y R, Xu Y, Wang J, Liu J F and Ma Z S
\href{https://doi.org/10.1103/PhysRevB.99.045420}
{2019 \textit{Phys. Rev. B} \textbf{99}, 045420}

\bibitem{Oriekhov}
Oriekhov D O, Gorbar E V and Gusynin V P 
\href{https://doi.org/10.1063/1.5078627}
{2018 \textit{Low Temp. Phys. } \textbf{44}, 1313–1324}

\bibitem{blochl1994}
Bl{\"o}chl P E
\href{https://doi.org/10.1103/PhysRevB.50.17953}
{1994 \textit{Phys. Rev. B} \textbf{50}, 17953}

\bibitem{perdew1996}
Perdew J P, Burke K, and Ernzerhof M
\href{https://doi.org/10.1103/PhysRevLett.77.3865}
{1996 \textit{Phys. Rev. Lett.} \textbf{77}, 3865}

\bibitem{aouail2021}
Aouail N, Belkaid M N, Oukebdane A and Tedjini M H
\href{https://doi.org/10.31349/RevMexFis.67.061003}
{2021 \textit{Rev. Mex. Fis.} \textbf{67}, 6}

\bibitem{guo2012}
Guo S D and Liu B G
\href{https://doi.org/10.1088/1674-1056/21/1/017101}
{2012 \textit{Chin. Phys. B} \textbf{21}, 017101}

\bibitem{GneitingCredit}
Unpublished private communications with Clemens Gneiting. 

\end{thebibliography}
\end{document}